\documentclass[
preprintnumbers,
preprint,
a4paper,
amsmath, 
amssymb, 
floatfix,
nofootinbib,
]{revtex4}

\usepackage{graphicx}

\begin{document}

\title{Time evolution of a Nambu-Goto string \\ coiling around a Kerr black hole}

\author{Hirotaka Yoshino${}^{1,2}$}
\author{Kousuke Tanaka${}^{1}$}

\affiliation{${}^1$Department of Physics, Osaka Metropolitan University, Osaka 558-8585, Japan}

\affiliation{${}^2$Nambu Yoichiro Institute of Theoretical and Experimental Physics (NITEP),
Osaka Metropolitan University, Osaka 558-8585, Japan}

\preprint{OCU-PHYS-627, AP-GR-213}

\date{April 22, 2026}

%
%
\begin{abstract}
The interaction between a Nambu-Goto string and a Kerr black hole gives one of the methods
of energy extraction from a rotating black hole. Although the properties of such processes have been well studied
for rigidly rotating strings, little is known for non-rigidly rotating strings.
In this paper, we study time evolution of a Nambu-Goto string on the equatorial plane
of a Kerr spacetime, which sticks on the horizon and extends to spatial infinity. 
The time evolution is studied by the series expansion with respect to $t$ and the numerical simulations,
which give reliable results for $t\lesssim 4M$ and $t\lesssim 38M$, respectively, where $M$ is the black hole mass.
Since the angular velocity of the string on the horizon
must coincide with the horizon angular velocity to keep the timelike property,
the string is dragged into rotation and coils around the black hole.  
The negative energy is observed to fall into the black hole, but the positive energy follows after that,
meaning that the energy extraction occurs for a short period of time.
In the outside region, a wave is generated and propagates to the distant region carrying the extracted energy.
After the propagation of the wave,
the system approaches the time-independent configuration found by
Boos and Frolov, and the total extracted energy is estimated as $E_{\rm ext}\lesssim \mu M$, 
where $\mu$ is the tension of the string.
\end{abstract}


\maketitle

%
%

\section{Introduction}
\label{Sec:Introduction}

There are three representative methods of energy extraction from rotating black holes:
The Penrose process \cite{Penrose:1971} (or the collisional and magnetic Penrose processes \cite{Schnittman:2018,Stuchlik:2021}), 
the Blandford-Znajek process \cite{Blandford:1977}, and
the superradiance or the superradiant instability (Ref.~\cite{Brito:2015} for a review).
These three processes have long research histories, and active researches are still ongoing motivated by the expectations
that these processes are actually occurring in our universe. 
Other than these three, 
there is a fourth method which make use of a Nambu-Goto string, which is a simplified model of a cosmic string. 
Although there exist early works on the phenomena of Nambu-Goto strings
in black hole spacetimes  (e.g., \cite{Lonsdale:1988,Frolov:1988,Frolov:1996-1,Frolov:1996-2,DeVilliers:1997,DeVilliers:1998,Frolov:1999,Snajdr:2002,Dubath:2006}), 
researches from the viewpoint of the energy extraction
appeared relatively recently (but see \cite{Frolov:2000} for a related earlier work). 
The first research that pointed out the possibility of energy extraction with a Nambu-Goto string
is Ref.~\cite{Kinoshita:2016}, where
the configurations of rigidly rotating Nambu-Goto strings were solved
up to the inner light surfaces.
More complete solutions of Nambu-Goto strings that stick to the horizons by crossing the inner light surfaces were constructed in Ref.~\cite{Igata:2018}.

Although the possibility that we can observe the interaction between a cosmic string and a black hole
may be small, this topic provides us with a physically interesting process at least theoretically. 
Also, it has been pointed out in Ref.~\cite{Kinoshita:2017} that the interaction between a Nambu-Goto string and a rotating black hole
can be regarded as a simplified process of the Blandford-Znajek process, because
the tension of magnetic field lines plays an important role in the Blandford-Znajek process while
a Nambu-Goto string has tension as well. In other words, since the two processes share the same feature, the 
study of a Nambu-Goto string helps us understand the physical interpretation
of the Blandford-Znajek process more deeply.

In this paper, we would like to focus less symmetric configuration of Nambu-Goto strings in dynamical situation.
Interesting simulations of a loop-shaped Nambu-Goto string was performed already in Ref.~\cite{Deng:2023}.
Here, we consider dynamical processes of an open Nambu-Goto string in the context of the energy extraction. 
In fact, we can expect interesting possibility from the analogy with the black hole magnetosphere.
In Refs.~\cite{Noda:2019,Noda:2021,Noda:2025}, the scattering of Alfv\'en waves was investigated
in the situation where the standard Blandford-Znajek process is occuring,
and there exist the cases where the reflected outgoing Alfv\'en waves
has a larger amplitude compared to the incident ingoing Alfv\'en waves.
These studies are interesting in that they provide a unified point of view of the Blandford-Znajek process and the superradiant scattering.
From this phenomenon, it is expected that waves generated on the rigidly rotating strings have possibility to extract energy 
from a rotating black hole as well.

Before studying scattering of waves on rigidly rotating strings, in this paper, 
we begin with a simpler situation as the first step.
We consider a Nambu-Goto string on the equatorial plane in a Kerr spacetime,
and suppose that it sticks to the horizon and extends to infinity.
More specifically, the string is supposed to be located on $\phi=0$ initially.
By the timelike property of the Nambu-Goto string, on the event horizon, the string must rotate in the $+\phi$
direction with the angular velocity that is identical to the horizon angular velocity.
Therefore, the string is dragged into rotation, and thus, it coils around the black hole
in time. Because the Nambu-Goto string has tension, it tends to pull back the rotation of the black hole,
and therefore, the energy extraction is expected to occur.
We would like to clarify whether this expectation is right or not using semianalytic and numerical methods.

This paper is organized as follows. In Sec.~\ref{Sec:Setup}, we explain the setup of the system
and the equation for the time evolution of a Nambu-Goto string.
In Sec.~\ref{Sec:methods}, we explain two methods for solving this equation,
the series expansion with respect to $t$ and the fully numerical computation.
The numerical results are presented in Sec.~\ref{Sec:Numerical-results}.
The properties of the obtained solutions and the amount of extracted energy are discussed.
Section~\ref{Sec:Summary} is devoted to a summary.
In Appendix~\ref{Appendix_A}, the formulas for the conserved quantities in this system, the energy and the angular momentum,
are derived. In Appendix~\ref{Appendix_B}, 
some of the detailed formulas for the method of series expansion are presented.

%
%
\section{Setup and the equation}
\label{Sec:Setup}

In this paper, we study the Nambu-Goto string moving in the equatorial plane of a Kerr spacetime.
The metric of the Kerr spacetime in the Boyer-Lindquist coordinates $x^\mu=(t,\phi, r, \theta)$ is
\begin{equation}
ds^2 \ = \ g_{tt}dt^2 + 2g_{t\phi}dtd\phi + g_{\phi\phi}d\phi^2 + g_{rr}dr^2 + g_{\theta\theta}d\theta^2,
\label{Kerr-metric}
\end{equation}
where the nonzero metric components are $g_{rr} = {\Sigma}/{\Delta}$, $g_{\theta\theta} =\Sigma$, and
\begin{subequations}
\begin{eqnarray}
g_{tt}&=&-\frac{\Delta - a^2\sin^2\theta}{\Sigma},\\
g_{t\phi} &=&-\frac{2Mar\sin^2\theta}{\Sigma},\\
g_{\phi\phi} &=& \frac{(r^2+a^2)^2-\Delta a^2\sin^2\theta}{\Sigma}
\sin^2\theta,
\end{eqnarray}
\end{subequations}
with
\begin{subequations}
\begin{eqnarray}
\Sigma &=& r^2+a^2\cos^2\theta, \\ 
\Delta &=& r^2+a^2-2Mr.
\end{eqnarray}
\end{subequations}
Here, $M$ is the Arnowitt-Deser-Misner (ADM) mass and $a$ is the rotation parameter with which
the gravitational angular momentum is given by $J=Ma$. 
The equation $\Delta=0$ has two solutions, $r=r_\pm$, with
\begin{equation}
r_{\pm} \ = \ M\pm\sqrt{M^2-a^2},
\end{equation}
and the larger solution $r=r_+$ corresponds to the event horizon.

The worldsheet of the Nambu-Goto string can be parametrized as
\begin{equation}
x^\mu  \ = \ x^\mu (\zeta^0, \zeta^1)
\end{equation}
using the two-dimensional coordinates $\zeta^a = (\zeta^0, \zeta^1)$ on the worldsheet.
The metric induced on the worldsheet can be written as
\begin{equation}
ds^2 \ = \ \sum_{a,b=0}^{1} \gamma_{ab}d\zeta^a d\zeta^b,
\end{equation}
and the action for the Nambu-Goto string is
\begin{equation}
S \ = \ \mu \int \sqrt{-\gamma}d\zeta^0d\zeta^1,
\label{action}
\end{equation}
where $\mu$ denotes the string tension and $\gamma$ is the determinant of the metric $\gamma_{ab}$.
There is a gauge degree of freedom for the 
coordinates $\zeta^a$. We fix this gauge degree of freedom 
by choosing
\begin{equation}
t \ = \ \zeta^0, \qquad r \ = \ \zeta^1.
\end{equation}
This means that the position of the Nambu-Goto string on the equatorial plane is specified as $\theta = \pi/2$ and
\begin{equation}
\phi = \Phi(t,r).
\end{equation}
Substituting $d\theta=0$ and $d\phi = \dot{\Phi}dt + \Phi^\prime dr$ into Eq.~\eqref{Kerr-metric},
where $\dot{\Phi}$ and $\Phi^\prime$ denotes the derivative of $\Phi(t,r)$ with respect to $t$ and $r$, respectively,
we have 
\begin{subequations}
\begin{eqnarray}
\gamma_{00} &= &  g_{tt}+2g_{t\phi}\dot{\Phi}+g_{\phi\phi}\dot{\Phi}^2,
\label{gamma00}
\\
\gamma_{01} &=& \Phi^\prime\left(g_{t\phi}+g_{\phi\phi}\dot{\Phi} \right),
\label{gamma01}
\\
\gamma_{11} &=& g_{rr}+g_{\phi\phi}\Phi^{\prime 2},
\label{gamma11}
\end{eqnarray}
\end{subequations}
and
\begin{equation}
-\gamma \, = \, 
\Delta\Phi^{\prime 2}-g_{rr}\left(g_{tt}+2g_{t\phi}\dot{\Phi}+g_{\phi\phi}\dot{\Phi}^2\right).
\label{Eq:determinant-gamma}
\end{equation}
Variation of the action in Eq.~\eqref{action} with respect to $\Phi$ gives the equation for $\Phi$
as
\begin{equation}
\frac{\partial}{\partial r}\left[\frac{\Delta\Phi^\prime}{\sqrt{-\gamma}}\right] 
\ = \ \frac{\partial}{\partial t}\left[\frac{g_{rr}\left(g_{t\phi}+g_{\phi\phi}\dot{\Phi}\right)}{\sqrt{-\gamma}}\right].
\end{equation}
This equation can be rewritten as
\begin{multline}
-\Delta\left(g_{rr}+g_{\phi\phi}\Phi^{\prime 2}\right)\ddot{\Phi}
-\Delta\left(g_{tt}+2g_{t\phi}\dot{\Phi}+g_{\phi\phi}\dot{\Phi}^2\right)
\left[\Phi^{\prime\prime}+\left(\frac{\Delta^\prime}{\Delta}-\frac{g_{rr}^\prime}{2g_{rr}}\right)\Phi^\prime\right]
\\
+2\Delta\left(g_{t\phi}+g_{\phi\phi}\dot{\Phi}\right)\Phi^\prime\dot{\Phi}^\prime
+\frac{\Delta\Delta^\prime}{2g_{rr}}\Phi^{\prime 3}
+\frac12\Delta\left(g^\prime_{tt}+2g^\prime_{t\phi}\dot{\Phi}+g^\prime_{\phi\phi}\dot{\Phi}^2\right)\Phi^\prime \, = \, 0.
\label{equation-for-string}
\end{multline}
In the textbook by Vilenkin and Shellard \cite{Vilenkin:2000}, the equation for the Nambu-Goto string that is applicable for an arbitrary gauge is given as
\begin{equation}
\frac{1}{\sqrt{-\gamma}}\partial_a\left(\sqrt{-\gamma}\gamma^{ab}x^\mu_{,b}\right)
+\Gamma^\mu_{\nu\sigma}\gamma^{ab}x^\nu_{,a}x^\sigma_{,b}\ =\ 0,
\end{equation}
and we have checked that all these equations agree with Eq.~\eqref{equation-for-string}.

We consider the situation where the Nambu-Goto string is initially located on the line $\phi=0$, i.e.,
\begin{equation}
\Phi(0,r)\ = \ 0.
\end{equation}
In this situation, one of the two ends of the Nambu-Goto string is sticking to the horizon while the other 
is located at spatial infinity. 
We consider the Nambu-Goto string which is timelike outside the event horizon, i.e., $-\gamma>0$. 
This requirement leads to the boundary condition for $\Phi(t,r)$ on the horizon. 
The value of $-\gamma$ in Eq.~\eqref{Eq:determinant-gamma} is rewritten as
\begin{equation}
-\gamma \ = \ \Delta \Phi^{\prime 2}+\frac{r^2}{g_{\phi\phi}}-\frac{r^2}{\Delta} g_{\phi\phi}
\left(\dot{\Phi}+\frac{g_{t\phi}}{g_{\phi\phi}}\right)^2,
\end{equation}
and assuming the finiteness of $\Phi^\prime$ on the horizon (which is confirmed by direct calculations of time evolution in the following section),
 one must impose
\begin{equation}
\dot{\Phi} \to -\frac{g_{t\phi}}{g_{\phi\phi}} \ = \ \Omega_{\rm H} \quad (r\to r_{+})
\end{equation}
in order to keep positivity of $-\gamma$ in the region $r\ge r_{+}$,
where $\Omega_{\rm H}$ is the angular velocity of the event horizon,
\begin{equation}
\Omega_{\rm H} \ = \ \frac{a}{2Mr_{+}}.
\end{equation}
Therefore, the Nambu-Goto string must rotate at the horizon angular velocity
on the event horizon, 
\begin{equation}
\dot{\Phi}(t,r_+) \ = \ \Omega_{\rm H},
\label{horizon-boundary-condition}
\end{equation}
and thus, the Nambu-Goto string coils around the black hole in time.

In this paper, as the case that guarantees the horizon boundary condition of Eq.~\eqref{horizon-boundary-condition}, 
we adopt
\begin{equation}
\dot{\Phi}(0, r) \ = \ \Omega_{\rm zamo}(r) := -\frac{g_{t\phi}}{g_{\phi\phi}}
\label{initial-condition-Phidot}
\end{equation}
for the initial condition of $\dot{\Phi}$. Here, $\Omega_{\rm zamo}(r)$ is the angular velocity
of the zero-angular-momentum-observers (ZAMOs), which stay at constant $r$ and $\theta$
while rotate in the $\phi$ direction so that their angular momenta are zero.
The conserved quantities, the energy and angular momentum, of the Nambu-Goto string are
calculated in Appendix~\ref{Appendix_A}, and the initial condition of Eq.~\eqref{initial-condition-Phidot}
together with Eq.~\eqref{dJdr}
implies that total angular momentum of the string is zero.

As for the boundary condition on the horizon, a more subtle issue of imposing the ingoing boundary condition
arises in the numerical simulation.
This point will be discussed in Sec.~\ref{Sec:numerical-method}.

%
%
\section{Two methods for solving the equation}
\label{Sec:methods}

In this section, we explain the methods for solving the motion of the Nambu-Goto string
in our setup. We apply two methods, the series expansion and the numerical simulation.

\subsection{Series expansion}
\label{Sec:series}

Since we have imposed $\Phi(0,r)=0$ as the initial condition, the Nambu-Goto string must have
the symmetry under the transformation, $t\to -t$ and $\phi\to -\phi$.
For this reason, $\Phi(t,r)$ becomes an odd function of $t$, and 
can be expanded as
\begin{equation}
\Phi(t,r) \ = \ \Omega_1(r)t + \frac{1}{3!}\Omega_3(r)t^3 + \frac{1}{5!}\Omega_5(r)t^5+\cdots,
\label{series-expansion}
\end{equation}
in the neighborhood of $t=0$. Since we have imposed $\dot{\Phi}(0,r) = \Omega_{\rm zamo}(r)$, 
the function $\Omega_1(r)$ in the series expansion is
\begin{equation}
\Omega_1(r) \ = \ \Omega_{\rm zamo}(r).
\end{equation}
We now substitute Eq.~\eqref{series-expansion} into the left-hand side of the equation for $\Phi$,
Eq.~\eqref{equation-for-string}. 
The left-hand side of Eq.~\eqref{equation-for-string} also becomes a polynomial function with odd powers of $t$,
and we require the coefficients of $t^n$ ($n=1,\, 3,\, 5,...$) to be zero. 
Then, we find that the coefficient of $t^n$ gives the equation for determining $\Omega_{n+2}$ in terms of
$\Omega_n$, $\Omega_{n-2}$, ..., $\Omega_1$ and their derivatives.
Schematically, they can be written as
\begin{subequations}
\begin{eqnarray}
\Omega_3(r)  &=& F_3(\Omega_1,\Omega_1^\prime,\Omega_1^{\prime\prime}), 
\label{F3}\\
\Omega_5(r)  &=& F_5(\Omega_1,\Omega_1^\prime,\Omega_1^{\prime\prime},\Omega_3,\Omega_3^\prime,\Omega_3^{\prime\prime}),
\label{F5}\\
\Omega_7(r)  &=& F_7(\Omega_1,\Omega_1^\prime,\Omega_1^{\prime\prime},\Omega_3,\Omega_3^\prime,\Omega_3^{\prime\prime},\Omega_5,\Omega_5^\prime,\Omega_5^{\prime\prime}),
\label{F7}\\
&\vdots&\notag
\end{eqnarray}
\end{subequations}
This means that functions $\Omega_n(r)$ can be determined sequentially.
The specific forms of $F_n$ are presented in Appendix~\ref{Appendix_B} for first few functions.
Using Mathematica, we have calculated $\Omega_n(r)$ up to $n=19$,
and then, the formulas of Eq.~\eqref{series-expansion} gives a fairly good approximate solution within the 
convergence radius, which has turned out to be $t\lesssim 4M$. 
The approximate solution obtained here provide us with a good benchmark test 
for the correctness of the numerical simulations.

\subsection{Numerical simulation}
\label{Sec:numerical-method}

The second method is the time evolution by numerical simulations. 
Instead of $r$, we adopt the tortoise coordinate $r_*$ defined by
\begin{equation}
\frac{dr_*}{dr}\ = \ \frac{r^2+a^2}{\Delta},
\end{equation}
or, explicitly,
\begin{equation}
r_*=r+\frac{2M}{r_+-r_-}
\left(r_+\log\left|\frac{r-r_+}{M}\right|-r_-\log\left|\frac{r-r_-}{M}\right|\right).
\end{equation}
Then, we evolve the variable $\psi$ introduced by
\begin{equation}
\psi \ = \ \Phi - \Omega_{\rm zamo}\,t.
\label{Def:psi}
\end{equation}
We derive the equation for $\psi(t,r_*)$. Substituting Eq.~\eqref{Def:psi} into Eq.~\eqref{equation-for-string},
we have
\begin{multline}
-\Delta\left[\frac{g_{rr}}{g_{\phi\phi}}+\left(\psi^{\prime}+\Omega^\prime t\right)^2\right]\ddot{\psi}
+\Delta\left(\frac{\Delta}{g_{\phi\phi}^2}-\dot{\psi}^2\right)\psi^{\prime\prime}
+2\Delta\dot{\psi}\left(\psi^{\prime}+\Omega^\prime t\right)\dot{\psi}^\prime
+\Delta\Omega^\prime\left(\psi^{\prime}+\Omega^\prime t\right)\dot{\psi}
\\
+\left\{\left[\frac{\Delta}{2}\left(\frac{g_{rr}^\prime}{g_{rr}}+\frac{g_{\phi\phi}^\prime}{g_{\phi\phi}}\right)-\Delta^\prime\right]
\left(\psi^{\prime}+\Omega^\prime t\right)
-\Delta\Omega^{\prime\prime}t\right\}\dot{\psi}^2
\\
+\frac{\Delta}{g_{\phi\phi}^2}
\left\{
\frac{\Delta^\prime g_{\phi\phi}}{2g_{rr}}\left(\psi^{\prime}+\Omega^\prime t\right)^3
+\frac12\left[\Delta\left(-\frac{g_{rr}^\prime}{g_{rr}}+\frac{g_{\phi\phi}^\prime}{g_{\phi\phi}}\right)+\Delta^\prime\right]\left(\psi^{\prime}+\Omega^\prime t\right)
+\Delta\Omega^{\prime\prime}t
\right\}
 \, = \, 0,
\end{multline}
where we have omitted the subscript ``zamo'' of $\Omega_{\rm zamo}$ for simplicity.
Then, we rewrite this equation in terms of $r_*$ using
\begin{equation}
\psi^\prime \, = \, \frac{r^2+a^2}{\Delta}\,\psi_{,r_*},
\end{equation}
\begin{equation}
\psi^{\prime\prime}\, = \, \frac{(r^2+a^2)^2}{\Delta^2} \psi_{,r_*r_*}
+\frac{1}{\Delta^2}\left[2r\Delta-(r^2+a^2)\Delta^\prime\right]\psi_{,r_*}.
\end{equation}
The result is
\begin{equation}
\ddot{\psi} \ = \ \mathcal{H}(\psi_{,r_*},\,\psi_{,r_*r_*},\,\dot{\psi},\,\dot{\psi}_{,r_*},\,r_*),
\label{equation-for-simulations}
\end{equation}
where
\begin{multline}
\mathcal{H}(\psi_{,r_*},\,\psi_{,r_*r_*},\,\dot{\psi},\,\dot{\psi}_{,r_*},\,r_*) \ = \ \\
C^{-1}\left[B_1 \dot{\psi}\dot{\psi}_{,r_*} 
+B_2 \psi_{,r_*r_*}
+B_3 \psi_{,r_*}
+B_4 \dot{\psi}^2\psi_{,r_*}
+B_5 \dot{\psi}^2
+B_6 \dot{\psi}
+B_7+B_8+B_9
\right],
\end{multline}
with
\begin{subequations}
\begin{equation}
B_1 \ = \ 2(r^2+a^2)D,
\end{equation}
\begin{equation}
B_2 \ = \ \left(\frac{\Delta}{g_{\phi\phi}^2}-\dot{\psi}^2\right)(r^2+a^2)^2,
\end{equation}
\begin{equation}
B_3  \ = \ \frac{\Delta}{g_{\phi\phi}^2}\left[2r\Delta-(r^2+a^2)\Delta^\prime\right],
\end{equation}
\begin{equation}
B_4\ = \ -\left\{\Delta\left[\frac{r^2-a^2}{r}-\frac{g_{\phi\phi}^\prime}{2g_{\phi\phi}}(r^2+a^2)\right]
+\frac12(r^2+a^2)\Delta^\prime\right\},
\end{equation}
\begin{equation}
B_5 \ = \ \Delta\left\{\left[\Delta\left(\frac{1}{r}+\frac{g_{\phi\phi}^\prime}{2g_{\phi\phi}}\right)-\frac32\Delta^\prime\right]\Omega^\prime 
-\Delta\Omega^{\prime\prime}\right\} t
\end{equation}
\begin{equation}
B_6 \ = \ \Delta\Omega^\prime D,
\end{equation}
\begin{equation}
B_7 \ = \ \frac{\Delta^\prime}{2g_{\phi\phi}r^2}D^3,
\end{equation}
\begin{equation}
B_8 \ = \ \frac{\Delta}{g_{\phi\phi}^2}\left[\Delta\left(-\frac{1}{r}+\frac{g_{\phi\phi}^\prime}{2g_{\phi\phi}}\right)+\Delta^\prime\right]D
\end{equation}
\begin{equation}
B_9 \ = \ \frac{\Delta^3}{g_{\phi\phi}^2}\Omega^{\prime\prime}t,
\end{equation}
\end{subequations}
\begin{equation}
C \ = \ \frac{\Delta r^2}{g_{\phi\phi}}+D^2,
\end{equation}
and 
\begin{equation}
D\ = \ (r^2+a^2)\psi_{,r_*}+\Delta\Omega^\prime t.
\end{equation}
We adopt the sixth-order finite differencing method in the spatial direction,
and integrate the equation using the fourth-order Runge-Kutta method in time direction
in the range $r_*^{\rm (in)}\le r_*\le r_*^{\rm (out)}$.
Typically, we adopt $r_*^{\rm (in)}/M=-200$ and $r_*^{\rm (out)}/M=500$.

The inner boundary condition must be discussed.
Taking the limit $\Delta\to 0$, Eq.~\eqref{equation-for-simulations} is reduced to
\begin{equation}
\Phi_{,r_*}^2\ddot{\Phi} - 2\Phi_{,r_*}\dot{\Phi}\dot{\Phi}_{,r_*} + \dot{\Phi}^2\Phi_{,r_*r_*}
+\frac{\Delta^\prime}{2(r_+^2+a^2)}
\left(\dot{\Phi}^2-\Phi_{,r_*}^2\right)\Phi_{,r_*}
\ = \ 0.
\end{equation}
Although this equation is highly nonlinear, it allows the following solutions:
\begin{equation}
\Phi = \Phi(t\pm r_*).
\end{equation}
For this reason, we adopt the standard ingoing boundary condition at the inner boundary $r_*=r_*^{\mathrm{(in)}}$.
At the outer boundary $r_*=r_*^{\rm (out)}$, we adopt the standard outgoing boundary condition.

The method explained here includes the problem of numerical instability.
In the simulation, there is an artificial unstable mode which grows exponentially in time,
and unfortunately, the growth rate becomes larger as the resolution is increased.
For example, the simulation crashes at $t\simeq 20M$ for the grid size $\Delta r_*/M=1.0$,
while it crashes at $t\simeq 5M$ for the grid size $\Delta r_*/M=0.1$.
In order to handle this unstable mode, we added the Kreiss-Oliger dissipation term.
More specifically, the equation is modified as 
\begin{subequations}
\begin{eqnarray}
\dot{\psi} &=& \varphi -\varepsilon(r) \psi_{,r_*r_*r_*r_*},\\
\dot{\varphi} &=& \mathcal{H}(\psi_{,r_*},\psi_{,r_*r_*},\varphi,\varphi_{,r_*},r_*) -\varepsilon(r) \varphi_{,r_*r_*r_*r_*},
\end{eqnarray}
\end{subequations}
with 
\begin{equation}
\varepsilon(r) \ = \ \varepsilon_0\,\exp\left[-\left(\frac{r_*}{100M}\right)^2\right],
\end{equation}
where $\varepsilon_0$ is adopted to be a value in the range $0.815\times 10^{-4}\le \varepsilon_0\le 0.825\times 10^{-4}$ depending on the situation.
This modification enables us to carry out the simulation up to $t\simeq 38M$. 
In this period of time, the Kreiss-Oliger dissipation term does not change the numerical solution significantly,
and its main role is just to suppress the growth of numerically unstable mode.

%
%
\section{Numerical results}
\label{Sec:Numerical-results}

We now explain the numerical results. First, we focus on the rapidly rotating case, $a/M=0.99$,  as the case where the effect of the energy extraction
is expected to be relevant.
After that, we briefly discuss the dependence of the results on $a/M$.

\subsection{Code check}

\begin{figure}[tb]
\centering
\includegraphics[width=0.60\textwidth]{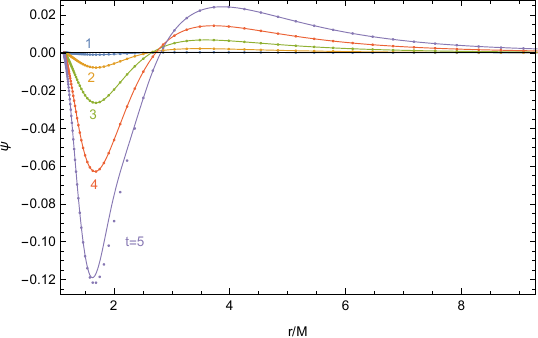}
\caption{Comparison of the values of $\psi(r)$ obtained by the series expansion (solid curves) and the numerical simulation (dots) for $t/M=1$, $2$, $3$, $4$, and $5$. The two results agree well for $t/M\lesssim 4$.}
\label{Fig:series-expansion4}
\end{figure}

As a code check, we compare the numerical solution with the approximate solution by the series expansion.
Figure~\ref{Fig:series-expansion4} shows the values of $\psi(t,r)$ for $t/M=0$, $1$, $2$, ..., $5$. Up to $t\lesssim 4M$, the two
solutions coincide remarkably well. A visible difference between the two solutions can be confirmed for $t/M=5$.
This would be because $t=5M$ is the out of the convergence radius of the series expansion.


%

\begin{figure}[tb]
\centering
\includegraphics[width=0.60\textwidth]{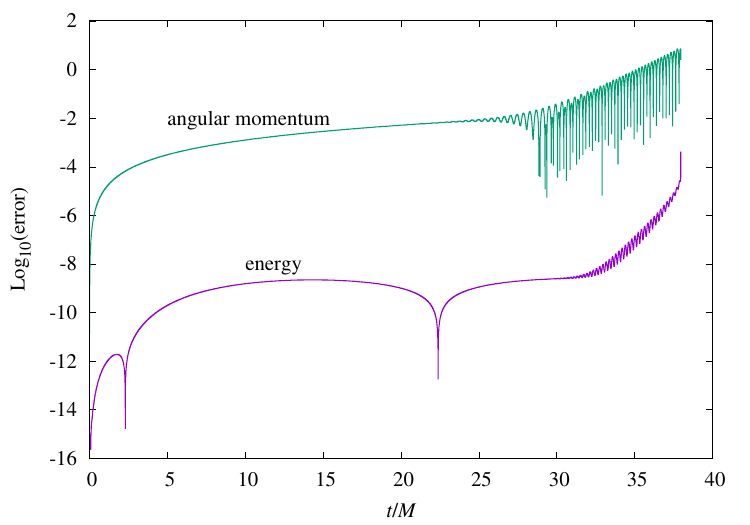}
\caption{Errors in the energy and the angular momentum defined in Eqs.~\eqref{error_E} and \eqref{error_J} as functions of time.}
\label{Fig:error}
\end{figure}
%

The conservation of the energy and the angular momentum is also monitored in the simulation.
The lower curve of Fig.~\ref{Fig:error} shows the absolute value of the difference in the energy from its initial value,
\begin{equation}
\mathrm{(error)}_E :=\left|\frac{E(t)-E(0)}{E(0)}\right|
\label{error_E}
\end{equation}
in the logarithmic scale
as a function of time in the compuation domain, $-200M\le r_*\le 500M$.
The error in the energy is $O(10^{-9})$ until $t/M\simeq 30$, after which
the growth of the artificial instability becomes relevant.
Although $\mathrm{(error)}_E$ is very small, we have to take care because the denominator 
of the right-hand side of Eq.~\eqref{error_E} becomes larger as the outer boundary is located at more distant position.
The right panel of the same figure shows the absolute value of the angular momentum,
\begin{equation}
\mathrm{(error)}_J :=\left|\frac{J(t)}{\mu M^2}\right|
\label{error_J}
\end{equation}
 in the logarithmic scale
as a function of time. Here, since the initial value of the angular momentum is zero, we do not normalize with the initial value.
Although $\mathrm{(error)}_J$ is larger compared to the energy, 
this would be because of the unnatural choice of the normalization factor.
The results here can be interpreted as a supportive evidence that the error is acceptable at least up to $t/M\simeq 35$.

\subsection{Properties of the solution and the energy extraction}


\begin{figure}[tb]
\centering
\includegraphics[width=0.95\textwidth]{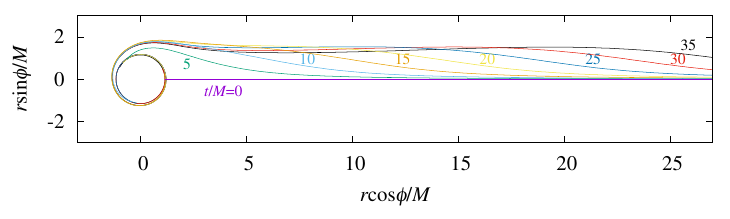}
\caption{The snapshots of the string positions for $t/M=0$, $5$, ..., $35$ on the equatorial plane
[i.e. $(r\cos\phi, r\sin\phi)$-plane].}
\label{Fig:equatorial-snapshots_paper}
\end{figure}
%

We now show the properties of the numerical solution. 
Figure~\ref{Fig:equatorial-snapshots_paper} shows the positions of the Nambu-Goto string on the equatorial plane
[i.e., $(r\cos\phi, r\sin\phi)$-plane] for $t/M=0$, $5$, ..., $35$.
Dragged by the rotation of the horizon, the Nambu-Goto string coils around the black hole.
During this process, a wave of the wavelength of $\sim 2\pi/\Omega_{\rm H}$ is generated
and it propagates to the distant region.


\begin{figure}[tb]
\centering
\includegraphics[width=0.45\textwidth]{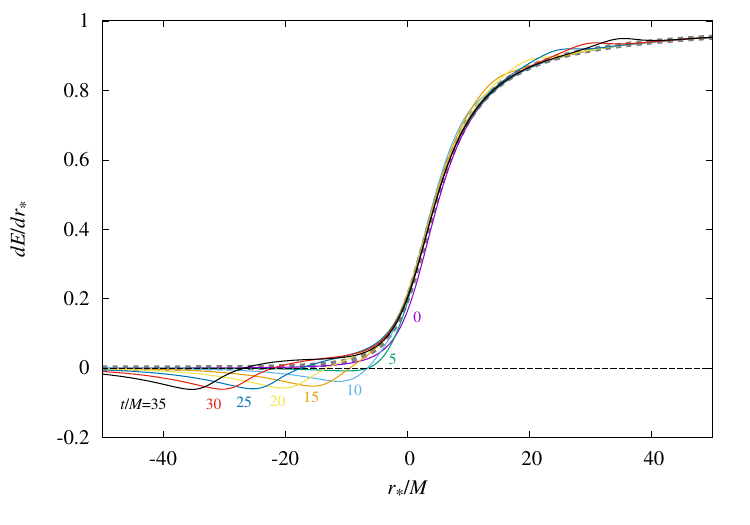}
\includegraphics[width=0.45\textwidth]{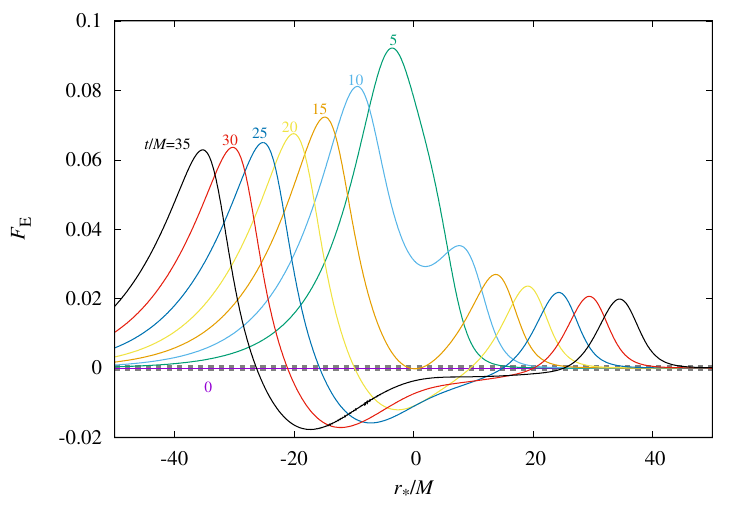}
\caption{The snapshots of the energy density with respect to the tortoise coordinate $dE/dr_*$ (the left panel) 
and the outgoing energy flux $F_E$ (the right panel) as functions of the tortoise coordinate $r_*/M$
for $t/M=0$, $5$, ..., $35$. 
The values of $dE/dr_*$ and $F_E$ are normalized with the string tension, $\mu$.
The values for the Boos-Frolov time-independent solution are indicated with dotted curves for comparison,
and in the left panel, the horizontal axis is shown with a thin dashed line.}
\label{Fig:E-density-flux}
\end{figure}
%

The left and right panels of Fig.~\ref{Fig:E-density-flux} show the snapshots of the energy density with
respect to the tortoise coordinate, $dE/dr_*$, and the outgoing energy flux crossing each $r=$constant surface, $F_E$,
as functions of $r_*/M$ for $t/M=0$, $5$, $10$, ..., $35$.
The necessary formulas for computing these quantities are presented in Appendix~\ref{Appendix_A}.
Here, $dE/dr_*$ and $F_E$ are normalized with the string tension, $\mu$. 
In the left panel, we can see a bump in the region with positive $r_*$ which propagates
in the outward direction, and correspondingly, there is also a bump in the right panel. 
This is the energy carried by the generated wave.
On the other hand, there is a local minimum of $dE/dr_*$ (left panel) in the region with negative $r_*$,
and the negative energy distributes around there. The position of the local minimum
shifts to the left-hand side in time, which means that the negative energy falls toward the horizon.
Correspondingly, there is a positive outgoing energy flux near the horizon
in the right panel, meaning that the extraction of energy takes place.
However, after that, a region with positive $dE/dr_*$ appears and that positive energy
falls toward the horizon following the negative energy. 
In the right panel, we can confirm the appearance of a region where $F_E$ becomes negative
due to the infall of the positive energy.
This means that the energy extraction stops after a short period of time.


\begin{figure}[tb]
\centering
\includegraphics[width=0.45\textwidth]{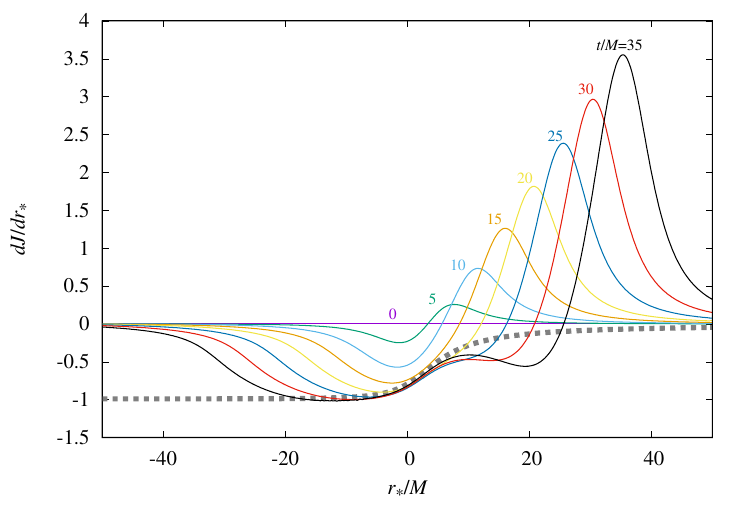}
\includegraphics[width=0.45\textwidth]{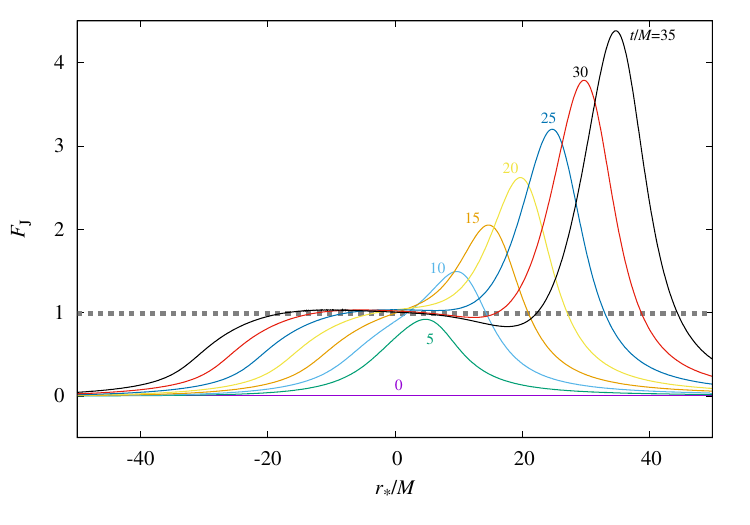}
\caption{The same as Fig.~\ref{Fig:E-density-flux} but for the angular momentum density $dJ/dr_*$ (the left panel) 
and the outgoing angular momentum flux $F_J$ (the right panel).
The values of $dJ/dr_*$ and $F_J$ are normalized with $\mu M$.}
\label{Fig:AM-density-flux}
\end{figure}
%

Figure~\ref{Fig:AM-density-flux} shows the snapshots of 
the angular momentum density with respect to the tortoise coordinate, $dJ/dr_*$ (the left panel), 
and the outgoing angular momentum flux, $F_J$ (the right panel)
as functions of $r_*/M$ for $t/M=0$, $5$, $10$, ..., $35$,
where the necessary formulas for computing these quantities are presented in Appendix~\ref{Appendix_A}.
There is a bump in the region $r_*/M>0$ whose position shifts to the right-hand side in time.
This means that the positive angular momentum is carried by the generated wave.
After the propagation of the wave, the region with negative $dJ/dr_*$ develops from the vicinity of the horizon
to the distant place. The value of outgoing angular momentum flux, $F_J$, is positive everywhere
all time. This means that the angular momentum continues to be extracted from the black hole throughout the simulation.

\subsection{Comparison with the Boos-Frolov solution}


\begin{figure}[tb]
\centering
\includegraphics[width=0.60\textwidth]{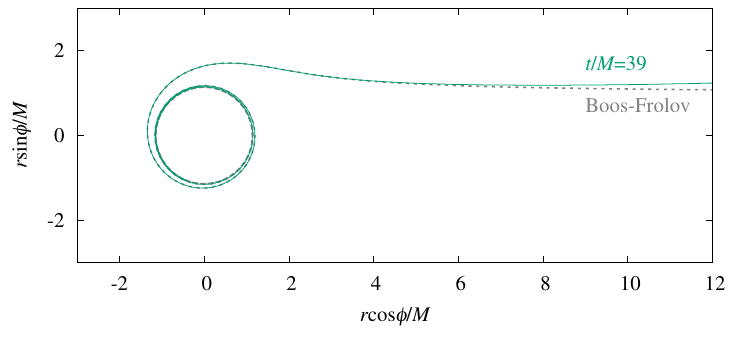}
\caption{The snapshots of the string positions for $t/M=39$ (the solid curve) and the Boos-Frolov solution (the dashed curve) on the equatorial plane.}
\label{Fig:compare-with-Boos-Frolov}
\end{figure}
%

At later time, the string has the tendency to approach the time-independent configuration
found by Boos and Frolov \cite{Boos:2017}.
The Boos-Frolov solution is given as
\begin{equation}
\Phi(t,r) \ = \ \frac{a}{r_+-r_-}\ln\left(\frac{r-r_+}{r-r_+}\right).
\end{equation}
Figure~\ref{Fig:compare-with-Boos-Frolov} shows the comparison of our numerical result at $t=39M$ and the Boos-Frolov solution.
Except for the neighborhood of the horizon and the distant place, the two solutions coincide well.
Therefore, it would be natural to expect that the present system asymptotes to the Boos-Frolov configuration
after the generated wave propagates to infinity. 
The values of $\mu^{-1}dE/dr_*$ and $\mu^{-1}dJ/dr_*$ calculated from Eqs.~\eqref{dEdr*} and \eqref{dJdr*}
for the Boos-Frolov solution,
\begin{subequations}
\begin{eqnarray}
\frac{1}{\mu}\frac{dE^{\rm (BF)}}{dr_*} &=& \frac{\Delta}{r^2+a^2},\\
\frac{1}{\mu}\frac{dJ^{\rm (BF)}}{dr_*} &=& -\frac{2Mar}{r^2+a^2},
\end{eqnarray}
\end{subequations}
are plotted by thick dotted curves in the left panels of Figs.~\ref{Fig:E-density-flux} and \ref{Fig:AM-density-flux}, respectively.
In the Boos-Frolov solution, the outgoing energy flux is zero everywhere from Eq.~\eqref{FE-r}, $F_E^{\rm (BF)}=0$, 
and the outgoing angular momentum flux is $F_J^{\rm (BF)}=\mu a$ everywhere from Eq.~\eqref{FJ-r}.
$F_E^{\rm (BF)}$ and $F_J^{\rm (BF)}$ are also plotted by thick dotted lines in the right panels
of Figs.~\ref{Fig:E-density-flux} and \ref{Fig:AM-density-flux}, respectively.
From $F_E^{\rm (BF)}=0$ and $F_J^{\rm (BF)}=\mu a$, if the system asymptotes to the Boos-Frolov state, 
the energy extraction will completely stop while
the angular momentum will continue to be extracted.
In order to confirm this expectation, a different method that enables a longer simulation must be developed.

\subsection{On the amounts of the extracted energy}
\label{Sec:E-amount}

\begin{figure}[tb]
\centering
\includegraphics[width=0.60\textwidth]{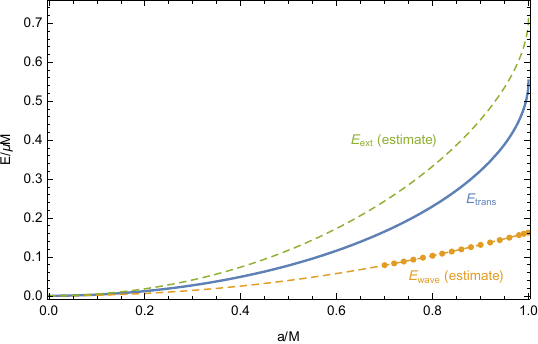}
\caption{The energy $E_{\rm trans}$ that is required to make transition from the initial state to the Boos-Frolov state
(the solid curve),
(the estimate of) the energy $E_{\rm wave}$ carried by the generated wave (the lower dashed curve with dots),
and (the estimate of) the total extracted energy $E_{\rm ext}$ from the black hole (the upper dashed curve),
as functions of $a/M$. Each energy is normalized with $\mu M$.}
\label{Fig:transition-energy}
\end{figure}

Here, we discuss the amounts of the extracted energy in this process taking account of the cases $a/M\neq 0.99$ as well.
We begin with pointing out the that the extracted energy can be divided into two parts. 
Comparing the Boos-Frolov state and the initial state of our system,
we find 
\begin{equation}
\left.\frac{dE}{dr}\right|_{t=0} \ = \ \frac{r^2}{\sqrt{(r^2+a^2)^2-\Delta a^2}}\frac{dE^{\rm (BF)}}{dr} \ < \ \frac{dE^{\rm (BF)}}{dr} \ = \ \mu,
\end{equation}
and thus, the Boos-Frolov state has larger energy compared to the initial state. 
Therefore, the transition from the initial state to the Boos-Frolov state requires the extraction of the energy from the black hole,
whose amount is denoted as 
\begin{equation}
E_{\rm trans} \ = \ E^{\rm (BF)} - E(0).
\end{equation}
The value of $E_{\rm trans}/\mu M$ is shown as a function of $a/M$ by a solid curve in Fig.~\ref{Fig:transition-energy}.
$E_{\rm trans}/\mu M$ is an increasing function of $a/M$, and is $O(1)$ for the near-extremal state.
In addition to $E_{\rm trans}$, there exists the energy carried by the generated wave, $E_{\rm wave}$.
The total amount of energy extracted from the black hole, $E_{\rm ext}$, 
must be
\begin{equation}
E_{\rm ext} \ = \ E_{\rm trans} + E_{\rm wave}.
\label{Eext}
\end{equation}
Unfortunately, the relaxation to the Boos-Frolov state is insufficient in our simulation
due to the limited duration of the simulation time, and 
we could not reproduce $E_{\rm trans}$ in our simulations.
For this reason, here we just evaluate $E_{\rm wave}$ by calculating the amount of the energy included in the bump 
in the left panel of Fig.~\ref{Fig:E-density-flux}. 
Physically, $E_{\rm wave}$ would mean the part of the extracted energy of which distant observers can make use.


%

In order to evaluate $E_{\rm wave}$, we calculate the difference in the energy
from that of the Boos-Frolov state in the region including the bump with
\begin{equation}
E_{\rm wave} \ = \ \int_{r_*^{\rm (L)}}^{r_*^{\rm (out)}} \left(\left.\frac{dE}{dr_*}\right|_{t/M=38} - \frac{dE^{\rm (BF)}}{dr_*}\right)dr_*.
\end{equation}
There is an arbitrariness in choosing the lower bound of the integration, $r_*^{\rm (L)}$, and we proceeded as follows.
In the inside region of the bump position, there is a region where ${dE}/{dr_*}$ (at $t/M=38$) becomes smaller than $dE^{\rm (BF)}/dr_*$, 
and we have chosen the value of $r_*^{(L)}$ at which ${dE}/{dr_*}=dE^{\rm (BF)}/dr_*$ holds.
Unfortunately, there is no strong physical ground for this choice, and comparing the results with other choices of $r_*^{(L)}$
and $t/M$, we have to admit that the estimates here may contain the error of $O(10\%)$.
The values of $E_{\rm wave}$ are numerically evaluated for the rotation parameters in the region $0.7\le a/M\le 0.999$,
and the result is presented in Fig.~\ref{Fig:transition-energy} with dots. 
The empirical formula, $E_{\rm wave}/\mu M\simeq 0.1625\times (a/M)^{2.05}$, guessed from the numerical result, is shown by a lower dashed curve.
$E_{\rm wave}/\mu M$ is also an increasing function of $a/M$, and is smaller compared to $E_{\rm trans}/\mu M$.

The total amount of extracted energy, $E_{\rm ext}$, is also estimated using Eq.~\eqref{Eext}, and is plotted
in Fig.~\ref{Fig:transition-energy} with an upper dashed curve. 
Even for the near-extremal case, the total extracted energy is smaller than $\mu M$.

%
%
\section{Summary}
\label{Sec:Summary}

In this paper, we have studied the time evolution of a Nambu-Goto string that sticks to the horizon
and extends to spatial infinity. Initially, the string is assumed to be located on $\phi=0$ and to have
the angular velocity identical to that of the ZAMOs. By the timelike property of the string, on the event horizon, 
the string must rotate with the same angular velocity as the horizon angular velocity, $\Omega_{\rm H}$.
As a result, the string is dragged into rotation near the black hole, and thus, it coils around the black hole
in time. During the time evolution, an outgoing wave with the wavelength of $\sim 2\pi/\Omega_{\rm H}$ is generated,
and it carries energy to the distant place.
The negative energy is observed to develop around the horizon and to fall toward the horizon.
Therefore, the energy extraction occurs in this system.
However, after a short period of time, the positive energy begins to fall following the negative energy,
meaning that the energy extraction stops. 
The system is observed to approach the time-independent configuration whose solution was found by
Boos and Frolov \cite{Boos:2017}, for which energy extraction does not occur while angular momentum continues to be extracted.

The amount of the extracted energy has been discussed in Sec.~\ref{Sec:E-amount}.
The extracted energy $E_{\rm ext}$ is divided into two parts as $E_{\rm ext}=E_{\rm trans}+E_{\rm wave}$, 
where  $E_{\rm trans}$ is the energy that is necessary for
making transition from the initial state to the Boos-Frolov state, and $E_{\rm wave}$ is the energy carried by the generated wave.
Both of $E_{\rm trans}$ and $E_{\rm wave}$ have the order of $\mu M$ for near-extremal black holes,
and $E_{\rm wave}$ is smaller compared to $E_{\rm trans}$. The total extracted energy
is less than $\mu M$, indicating that the system studied in this paper is not very effective as the method of 
energy extraction from a rotating black hole.

Unfortunately, the code developed in this paper could not 
realize long-term simulations because of the growth of artificial numerical instability.
It is desirable to develop a code based on a different method that 
enables long-term simulations.
Developing such a code is also useful and necessary for studying more interesting situations.
As stated in Sec.~\ref{Sec:Introduction}, from the analogy with the combination of the superradiance
and the Blandford-Znajek process in black hole magnetosphere \cite{Noda:2019,Noda:2021,Noda:2025},
it is expected that waves generated on rigidly rotating strings in black hole spacetimes would also be useful as the method
of energy extraction from a rotating black hole. In this sense, the study of this paper is a starting
point for these interesting directions, and we hope to tackle these problems in future.

%
%
\acknowledgments

H.~Y. thanks Chul-moon Yoo for helpful comments and we also
thank colleagues of the astrophysics group at Osaka Metropolitan University.
H.~Y. is in part supported by JSPS KAKENHI Grant Numbers JP22H01220 and
JP21H05189,
and is partly supported by MEXT Promotion of Distinctive Joint Research Center Program  JPMXP0723833165.

\appendix


\section{Conserved quantities}
\label{Appendix_A}

If the spacetime possesses a Killing vector $\xi^\mu$, the conserved four-current for matter fields can be defined as
\begin{equation}
p^\mu \ = \ \mp T^{\mu\nu}\xi_\nu,
\label{four-current}
\end{equation}
where we choose minus and plus signs when $\xi^\mu$ is (asymptotically) timelike and spacelike, respectively.
The four-current $p^\mu$ satisfies the conserved law, $\nabla_\mu p^\mu = 0$, which is rewritten as
\begin{equation}
\frac{1}{\sqrt{-g}}\partial_\mu\left(\sqrt{-g}p^\mu\right) \ = \ 0.
\end{equation}
Integrating this equation by parts in the domain $\mathcal{D}$ given by
$t_1\le t\le t_2$, $r_+\le r \le r_{\rm out}$, and the whole angular domain of $(\theta,\phi)$, we have the conservation law,
\begin{equation}
C_\xi(t_2)
\ = \ 
C_\xi(t_1) - \int_{t_1}^{t_2}F_C(t,r_{\rm out})dt
+ \int_{t_1}^{t_2}F_C(t,r_{+})dt,
\end{equation}
where the conserved quantity on the spacelike hypersurface $t=\mathrm{constant}$ is defined as
\begin{equation}
C_\xi(t) = \int_{t} drd\theta d\phi \sqrt{-g}p^t,
\end{equation}
and the outward flux of the conserved current that crosses 
$r=\mathrm{constant}$ surface per unit coordinate time is defined as
\begin{equation}
F_C(t,r) = \int_{r} d\theta d\phi \sqrt{-g}p^r.
\end{equation}

The purpose of this Appendix is to derive the formulas for the energy and the angular momentum
in our setup. By varying the action of Eq.~\eqref{action} with respect to $g_{\mu\nu}$, 
the energy-momentum tensor of the Nambu-Goto string is derived as
(see p.~156 of Ref.~\cite{Vilenkin:2000})\footnote{Note that a minus sign is added to the formula of Ref.~\cite{Vilenkin:2000} because of the difference
of the sign convention for the metric.}
\begin{equation}
\sqrt{-g}\,T^{\mu\nu} \ = \ -\mu\int d^2\zeta \sqrt{-\gamma}\gamma^{ab}x^\mu_{,a}x^\nu_{,b}\delta^{(4)}(x^\sigma-x^\sigma(\zeta^a)).
\label{energy-momentum-tensor}
\end{equation}
For this energy-momentum tensor, we calculate the formulas for the energy and angular momentum, one by one.

\subsection{Energy}

The energy is the conserved quantity associated with the Killing vector $\xi_\nu = (\partial_t)_\nu$
(i.e., the stationarity of the spacetime), 
and we adopt the minus sign 
in Eq.~\eqref{four-current}. We calculate the energy density with respect to $r$ and $r_*$, and the energy flux.

\subsubsection{Energy density with respect to $r$ and $r_*$}

The energy in the Kerr spacetime is given by
\begin{equation}
E(t) \ = \ -\int_{t} drd\theta d\phi \sqrt{-g}\left(T^{tt}g_{tt}+T^{t\phi}g_{t\phi}\right).
\end{equation}
Substituting the $tt$ and $t\phi$ components of the energy momentum tensor of Eq.~\eqref{energy-momentum-tensor}
and carrying out the integration with respect to $\zeta^0$, $\zeta^1$, $\theta$, and $\phi$ (remember that in our gauge, $t=\zeta^0$,
$r=\zeta^1$, $\theta=\pi/2$, and $\phi=\Phi(\zeta^0, \zeta^1)$), we have
\begin{eqnarray}
{E(t)} 
\ = \ 
\mu\int_{t} dr\sqrt{-\gamma}\left[g_{tt}
\gamma^{00}
+g_{t\phi}\left(\gamma^{00}\Phi_{,0}+\gamma^{01}\Phi_{,1}\right)\right].
\end{eqnarray}
Rewriting $\gamma^{ab}$ using Eqs.~\eqref{gamma00}--\eqref{gamma11}, we obtain
\begin{eqnarray}
\frac{1}{\mu}\frac{dE}{dr} 
& = & \frac{1}{\sqrt{-\gamma}}\left[{\Delta\Phi^{\prime 2}-g_{rr}\left(g_{tt}+g_{t\phi}\dot{\Phi}\right)}\right].
\label{dEdr}
\end{eqnarray}
The energy density with respect to the tortoise coordinate $r_*$ is expressed as
\begin{equation}
\frac{1}{\mu}\frac{dE}{dr_*} \ = \ \frac{\sqrt{\Delta}}{r^2+a^2}
\frac{\left[(r^2+a^2)\psi_{,r_*}+\Delta\Omega_{\rm zamo}^\prime t\right]^2+r^2\Delta/g_{\phi\phi}-r^2g_{t\phi}\dot{\psi}}
{\sqrt{\left[(r^2+a^2)\psi_{,r_*}+\Delta\Omega_{\rm zamo}^\prime t\right]^2+r^2\Delta/g_{\phi\phi}-r^2g_{\phi\phi}\dot{\psi}^2}},
\label{dEdr*}
\end{equation}
where we have used Eq.~\eqref{Def:psi}.

\subsubsection{Energy flux with respect to the coordinate time}

Next, we calculate the outgoing energy flux $F_E(t,r)$ crossing an $r=$constant surface. The energy flux
is given by
\begin{equation}
F_E 
\ = \ 
-\int_{r} drd\theta d\phi \sqrt{-g}\left(T^{rt}g_{tt}+T^{r\phi}g_{t\phi}\right).
\end{equation}
Substituting $rt$ and $r\phi$ components of the energy momentum tensor of Eq.~\eqref{energy-momentum-tensor}
and carrying out the integration with respect to $\zeta^0$, $\zeta^1$, $\theta$, and $\phi$, we have
\begin{eqnarray}
F_E (t,r)
\ = \ 
 \mu\sqrt{-\gamma}\left[g_{tt}
\gamma^{10}
+g_{t\phi}\left(\gamma^{10}\Phi_{,0}+\gamma^{11}\Phi_{,1}\right)\right].
\end{eqnarray}
Rewriting $\gamma^{ab}$ using Eqs.~\eqref{gamma00}--\eqref{gamma11}, we obtain
\begin{eqnarray}
\frac{F_E}{\mu} & = & -\frac{\Delta}{\sqrt{-\gamma}}\Phi^\prime\dot{\Phi}.
\label{FE-r}
\end{eqnarray}
In terms of $\psi$ introduced in Eq.~\eqref{Def:psi} and the tortoise coordinate $r_*$, $F_E$ is expressed as
\begin{equation}
\frac{F_E}{\mu} \ = \ 
-\frac{\sqrt{\Delta}\,\left[(r^2+a^2)\psi_{,r_*}+\Delta\Omega^\prime t\right]\left(\dot{\psi}+\Omega\right)}
{\sqrt{\left[(r^2+a^2)\psi_{,r_*}+\Delta\Omega^\prime t\right]^2+r^2\Delta/g_{\phi\phi}-r^2g_{\phi\phi}\dot{\psi}^2}}.
\label{FE-r*}
\end{equation}

\subsection{Angular momentum}

The angular momentum is the conserved quantity associated with the Killing vector $\xi_\nu = (\partial_\phi)_\nu$
(i.e., the axial symmetry of the spacetime), 
and we adopt the plus sign in Eq.~\eqref{four-current}. 
We calculate the angular momentum density with respect to $r$ and $r_*$, and the angular momentum flux.

\subsubsection{Angular momentum density with respect to $r$ and $r_*$}

The angular momentum in the Kerr spacetime is given by
\begin{equation}
J(t) \ = \ \int_{t} drd\theta d\phi \sqrt{-g}\left(T^{tt}g_{t\phi}+T^{t\phi}g_{\phi\phi}\right).
\end{equation}
Substituting the $tt$ and $t\phi$ components of the energy momentum tensor of Eq.~\eqref{energy-momentum-tensor}
and carrying out the integration with respect to $\zeta^0$, $\zeta^1$, $\theta$, and $\phi$, we have
\begin{eqnarray}
J(t) & =&  -\mu\int_{t} dr \sqrt{-\gamma}\left[g_{t\phi}
\gamma^{00}
+g_{\phi\phi}\left(\gamma^{00}\Phi_{,0}+\gamma^{01}\Phi_{,1}\right)\right].
\end{eqnarray}
Rewriting $\gamma^{ab}$ using Eqs.~\eqref{gamma00}--\eqref{gamma11}, we obtain
\begin{equation}
\frac{1}{\mu}\frac{dJ}{dr} \ = \ \frac{1}{\sqrt{-\gamma}}{g_{rr}\left(g_{t\phi}+g_{\phi\phi}\dot{\Phi}\right)}.
\label{dJdr}
\end{equation}
The angular momentum density with respect to the tortoise coordinate $r_*$ is expressed as
\begin{equation}
\frac{1}{\mu}\frac{dJ}{dr_*} \ = \ \frac{\sqrt{\Delta}}{r^2+a^2}
\frac{r^2g_{\phi\phi}\dot{\psi}}
{\sqrt{\left[(r^2+a^2)\psi_{,r_*}+\Delta\Omega^\prime t\right]^2+r^2\Delta/g_{\phi\phi}-r^2g_{\phi\phi}\dot{\psi}^2}}.
\label{dJdr*}
\end{equation}
where we have used Eq.~\eqref{Def:psi}.

\subsubsection{Angular momentum flux with respect to the coordinate time}

Next, we calculate the outgoing angular momentum flux $F_J(t,r)$ crossing an $r=$constant surface. The angular momentum flux
is given by
\begin{equation}
F_J
\ = \ \int_{r} d\theta d\phi \sqrt{-g}\left(T^{rt}g_{t\phi}+T^{r\phi}g_{\phi\phi}\right).
\end{equation}
Substituting $rt$ and $r\phi$ components of the energy momentum tensor of Eq.~\eqref{energy-momentum-tensor}
and carrying out the integration with respect to $\zeta^0$, $\zeta^1$, $\theta$, and $\phi$, we have
\begin{eqnarray}
F_J (t,r)
\ = \ 
-\mu \sqrt{-\gamma} \left[g_{t\phi}
\gamma^{10}
+g_{\phi\phi}\left(\gamma^{10}\Phi_{,0}+\gamma^{11}\Phi_{,1}\right)\right].
\end{eqnarray}
Rewriting $\gamma^{ab}$ using Eqs.~\eqref{gamma00}--\eqref{gamma11}, we obtain
\begin{eqnarray}
\frac{F_J}{\mu} & = & -\frac{1}{\sqrt{-\gamma}}\Delta\Phi^\prime.
\label{FJ-r}
\end{eqnarray}
In terms of $\psi$ introduced in Eq.~\eqref{Def:psi} and the tortoise coordinate $r_*$, $F_J$ is expressed as
\begin{equation}
\frac{F_J}{\mu} \ = \ 
-\frac{\sqrt{\Delta}\,\left[(r^2+a^2)\psi_{,r_*}+\Delta\Omega^\prime t\right]}
{\sqrt{\left[(r^2+a^2)\psi_{,r_*}+\Delta\Omega^\prime t\right]^2+r^2\Delta/g_{\phi\phi}-r^2g_{\phi\phi}\dot{\psi}^2}}.
\label{FJ-r*}
\end{equation}


\section{Detailed formulas for the series expansion}
\label{Appendix_B}

In Sec.~\ref{Sec:series}, the method for the series expansion is explained, and the 
schematic formulas for determining $\Omega_{3}$, $\Omega_5$, $\Omega_7$,... are
presented in Eqs.~\eqref{F3}--\eqref{F7}. Here, we present the explicit forms of $F_3$, $F_5$, and $F_7$.
The function $F_3(\Omega_1, \Omega_1^\prime, \Omega_1^{\prime\prime})$ is given as
\begin{multline}
F_3 \ = \ \frac{2(g_{t\phi}+g_{\phi\phi}\Omega_1)}{g_{rr}}\Omega_1^{\prime 2}
+\frac{g_{tt}^\prime+2g_{t\phi}^\prime\Omega_1+g_{\phi\phi}^\prime\Omega_1^2}{2g_{rr}}\,\Omega_1^\prime
\\
-\frac{g_{tt}+2g_{t\phi}\Omega_1+g_{\phi\phi}\Omega_1^2}{g_{rr}}
\left[\Omega_1^{\prime\prime}-\left(\frac{g_{rr}^\prime}{2g_{rr}}-\frac{\Delta^\prime}{\Delta}\right)\Omega_1^\prime\right].
\label{F3-explicit}
\end{multline}
Recalling the fact that we have chosen $\Omega_1$ as $\Omega_1=-g_{t\phi}/g_{\phi\phi}$ in Eq.~\eqref{initial-condition-Phidot},
the first term of Eq.~\eqref{F3-explicit} can be eliminated.
The formulas for $F_5$ and $F_7$ are 
\begin{subequations}
\begin{multline}
F_5 \ = \ 
\frac{2(g_{t\phi}+g_{\phi\phi}\Omega_1)}{g_{rr}}
\left[4\Omega_1^\prime\Omega_3^\prime-3\Omega_3\Omega_1^{\prime\prime}
+3\left(\frac{g_{rr}^\prime}{2g_{rr}}-\frac{\Delta^\prime}{\Delta}\right)\Omega_3\Omega_1^\prime
\right]
\\
-\frac{g_{tt}+2g_{t\phi}\Omega_1+g_{\phi\phi}\Omega_1^2}{g_{rr}}\, \left[\Omega_3^{\prime\prime}-\left(\frac{g_{rr}^\prime}{2g_{rr}}-\frac{\Delta^\prime}{\Delta}\right)\Omega_3^\prime\right]
+\frac{g_{tt}^\prime+2g_{t\phi}^\prime\Omega_1+g_{\phi\phi}^\prime\Omega_1^2}{2g_{rr}}\, \Omega_3^\prime
\\
+\frac{3(g_{t\phi}^\prime+g_{\phi\phi}^\prime\Omega_1)}{g_{rr}}\,\Omega_3\Omega_1^\prime
+\frac{3\Delta^\prime}{g_{rr}^2}\,\Omega_1^{\prime 3},
\end{multline}
\begin{multline}
F_7 \ = \ 
\frac{2(g_{t\phi}+g_{\phi\phi}\Omega_1)}{g_{rr}}
\left[10\Omega_3^{\prime 2}-10\Omega_3\Omega_3^{\prime\prime}+6\Omega_1^\prime\Omega_5^\prime-5\Omega_5\Omega_1^{\prime\prime}
+5\left(\frac{g_{rr}^\prime}{2g_{rr}}-\frac{\Delta^\prime}{\Delta}\right)\left(\Omega_5\Omega_1^\prime+2\Omega_3\Omega_3^\prime\right)
\right]
\\
-\frac{10g_{\phi\phi}}{g_{rr}}
\left[\Omega_5\Omega_1^{\prime 2}-4\Omega_3\Omega_1^\prime\Omega_3^\prime+3\Omega_3^2\Omega_1^{\prime\prime}
-3\left(\frac{g_{rr}^\prime}{2g_{rr}}-\frac{\Delta^\prime}{\Delta}\right)\Omega_3^2\Omega_1^\prime
\right]
\\
-\frac{g_{tt}+2g_{t\phi}\Omega_1+g_{\phi\phi}\Omega_1^2}{g_{rr}}
\left[ \Omega_5^{\prime\prime} -\left(\frac{g_{rr}^\prime}{2g_{rr}}-\frac{\Delta^\prime}{\Delta}\right)\Omega_5^\prime\right]
+\frac{g_{tt}^\prime+2g_{t\phi}^\prime\Omega_1+g_{\phi\phi}^\prime\Omega_1^2}{2g_{rr}}\, \Omega_5^\prime
\\
+\frac{5(g_{t\phi}^\prime+g_{\phi\phi}^\prime\Omega_1)}{g_{rr}}\left(\Omega_5\Omega_1^\prime+2\Omega_3\Omega_3^\prime\right)
+\frac{15g_{\phi\phi}^\prime }{g_{rr}}\,\Omega_3^2\Omega_1^{\prime}
+\frac{30\Delta^\prime}{g_{rr}^2}\,\Omega_1^{\prime 2}\Omega_3^\prime.
\end{multline}
\end{subequations}
Again, the first terms of these equations can be omitted for the choice $\Omega_1=-g_{t\phi}/g_{\phi\phi}$.



\end{document}